\documentstyle[sprocl]{article}

\def\be{\begin{equation}}
\def\ee{\end{equation}}
\def\bea{\begin{eqnarray}}
\def\eea{\end{eqnarray}}

\begin{document}

\title{EINSTEIN-DE BROGLIE RELATIONS ON THE LATTICE}

\author{ M. LORENTE}

\address{Department or Physics, University of Oviedo,\\ 33007 Oviedo, Spain}

\maketitle\abstracts{ Historically the starting point of wave mechanics is the Planck and
Einstein-de Broglie relations for the energy and momentum of a particle, where the momentum
is connected to the group velocity of the wave packet. We translate the arguments given by
de Broglie to the case of a wave defined on the grid points of a space-time lattice and
explore the physical consequences such as integral period, wave length, discrete energy,
momentum and rest mass.}
  
\section{Einstein-de Broglie relations: continuous case}

After Einstein applied the Planck formula $E=h\nu$ (quantization of the energy for the
orbits of the harmonic oscillator) to the energy of the light waves in the photo-electric
efect, de Broglie generalized this expression to relativistic momentum of a massive or
massless particle.

Basicaly de Broglie arguments [1] are bases on the transformtions properties of the
frequency and wave number of a plane wave and the transformations of the energy and the
relativistic momentum of a particle.

Let us consider a plane wave with the wave normal $\vec{n}$ in the xy-plane of a system $S$
with angular frequency $w$, wave vector $\vec{k}$ and phase velocity $v_\varphi$. It is
described by a wave function
\begin{equation}
\psi (\vec{x},t)=A\ \cos\left({wt-\vec{k}\vec{x}}\right)
\end{equation}

In a coordinate system $S'$ moving in the direction of the x-axis with the velocity $v$
relative to $S$ the wave function will be described by
\begin{equation}
\psi \left({\vec{x}',t'}\right)=A\ \cos\left({w't'-\vec{k}'\cdot \vec{x}'}\right)
\end{equation}

Since the argument of both functions should be the same it follows: by elementary
calculations:
\[w'=w{\frac{\left({1-\vec{v}\cdot \vec{n}}\right)/{v}_{\varphi
}}{{\left({1-{v}^{2}/{c}^{2}}\right)}^{1/2}}}\]
\[\vec{k}'=\vec{k}+{\frac{\vec{v}}{{v}^{2}}}{\frac{\left({\vec{v}\cdot
\vec{k}}\right)\left\{{{1-\left({1-{v}^{2}/{c}^{2}}\right)}^{1/2}}\right\}-{v}^{2}k{v}_{\varphi
}/{c}^{2}}{{\left({1-{v}^{2}/{c}^{2}}\right)}^{1/2}}}\]

\begin{equation}
k'=k{\frac{{\left({1-{\frac{{v}^{2}}{{c}^{2}}}+{\frac{{v}^{2}{v}_{\varphi
}^{2}}{{c}^{4}}}+{\frac{{\left({\vec{v}\cdot
\vec{n}}\right)}^{2}}{{c}^{2}}}-{\frac{{2\left({\vec{v}\cdot \vec{n}}\right){v}_{\varphi
}}^{}}{{c}^{2}}}}\right)}^{1/2}}{{\left({1-{v}^{2}/{c}^{2}}\right)}^{1/2}}}
\end{equation}
where $k\equiv \left|{\vec{k}}\right|$ is the wave number.

Suppose a particle of energy $E$ and relativistic momentum $\vec{p}$ is moving with
respecto to a coordinate system $S$ with velocity $\vec{u}$. In a coordinate system $S'$
moving in the direction of the x-axis with the velocity $v$ relative to $S$, the particle
will be described by the energy $E'$ and the relativistic momentum $\vec{p'}$, which are
related to the old coordinates by
\[E'=E{\frac{1-\vec{v}\cdot \vec{u}/{c}^{2}}{{\left({1-{v}^{2}/{c}^{2}}\right)}^{1/2}}}\]
\[\vec{p}'=\vec{p}+{\frac{\vec{v}}{{v}^{2}}}{\frac{\left({\vec{v}\cdot
\vec{p}}\right)\left\{{{1-\left({1-{v}^{2}/{c}^{2}}\right)}^{1/2}}\right\}-{v}^{2}p/u}{{\left({1-{v}^{2}/{c}^{2}}\right)}^{1/2}}}\]

\begin{equation}
p'={\frac{{\left\{{{p}^{2}\left({1-{\frac{{v}^{2}}{{c}^{2}}}}\right)+{p}^{2}{\frac{{v}^{2}}{{u}^{2}}}+{\frac{{\left({\vec{v}\cdot
\vec{p}}\right)}^{2}}{{c}^{2}}}-{\frac{2p\left({\vec{v}\cdot
\vec{p}}\right)}{u}}}\right\}}^{1/2}}{{\left({1-{v}^{2}/{c}^{2}}\right)}^{1/2}}}
\end{equation}

Comparison of formulas (3) and (4) leads to the conclusion that $w,\vec{k}$ transform in
the same way as $E,\vec{p}$ provided $\vec{k}$ and $\vec{p}$ are parallel and the phase
velocity $v_\varphi$ is related to the velocity of the particle $u$ by the expression [2]
\begin{equation}
{v}_{\varphi }={c}^{2}/u
\end{equation}

Following Einstein's hypothesis that the energy should be proportional to the frecuency of
a light quanta,
\begin{equation}
E=\hbar w
\end{equation}
de Broglie made the assumption that for a particle there is an associate wave satisfying
\begin{equation}
E=\hbar w\ \ \ \ ,\ \ \ \ p=\hbar \vec{k}
\end{equation}

Since the phase velocity of the wave ${v}_{\varphi }$ does not correspond to the velocity
of the particle, de Broglie suggested that there is a wave packet associated with the
particle, consisting of a superposition of waves with different wave vectors $\vec{k}$ and
amplitudes $\hat{\psi }\left({\vec{k}}\right)$
\begin{equation}
\psi \left({\vec{x},t}\right)=\int_{-\infty }^{\infty
}{d}^{3}k\hat{\psi}\left({\vec{k}}\right)\exp\
i\left\{{w\left({\vec{k}}\right)t-\vec{k}\cdot
\vec{x}}\right\}
\end{equation}

If we suppose that the momentum vari is very little around a fixed value
$\vec{{k}_{0}}$, namely, $\left|{\vec{k}-\vec{{k}_{0}}}\right|\le \Delta k$, then the
function $w\left({\vec{k}}\right)$ can be expanded around ${w}_{0}\equiv
w\left({\vec{{k}_{0}}}\right)$. Easy calculations gives:
\[\psi \left({\vec{x},t}\right)=\exp\ \left\{{i\left({{w}_{0}t-\vec{{k}_{0}}\cdot
\vec{x}}\right)\Delta k}\right\}\int_{\Delta k}^{}{d}^{3}k\ \exp\
\left\{{i\left({{w'}_{0}t-x}\right)\Delta k}\right\}\]

This wave represent a packet with phase velocity ${v}_{\varphi }={w}_{0}/{k}_{0}$ and
group velocity ${v}_{g}={w'}_{0}\equiv dw/dk\left({{k}_{0}}\right)$.

From the Einstein-de Broglie relations follows:
\begin{equation}
{v}_{\varphi }={\frac{w}{k}}={\frac{E}{p}}={\frac{{c}^{2}}{u}}
\end{equation}
\begin{equation}
{v}_{g}={\frac{{\rm d}\,w}{{\rm d}\,k}}={\frac{{\rm d}\,E}{{\rm d}
\,p}}={\frac{{pc}^{2}}{E}}=u
\end{equation}
where in the last equation we have
used $E={\left({{p}^{2}{c}^{2}+{m}_{0}^{2}{c}^{4}}\right)}^{1/2}$. There fore we have
${v}_{\varphi }={c}^{2}/{v}_{g}$ in agreement with de Broglie assumption about the
wave-packet.

The Einstein-de Broglie relation were used to write the wave function associated to a
particle
\begin{equation}
\psi \left({\vec{x},t}\right)=\exp\ \left\{{i\left({Et-\vec{p}\cdot \vec{x}}\right)/\hbar
}\right\}
\end{equation}

If the energy and relativistic momentum are connected by
${E}^{2}-{p}^{2}{c}^{2}={m}_{0}^{2}{c}^{4}$ the wave function satisfies
\begin{equation}
\left({-{\frac{1}{{c}^{2}}}{\frac{\partial \,}{\partial \,{t}^{2}}}+\Delta }\right)\psi
\left({\vec{x},t}\right)={\frac{{m}_{0}^{2}{c}^{4}}{{\hbar }^{2}}}\psi
\left({\vec{x},t}\right)
\end{equation}

\section{Einstein-de Broglie relations: discrete case}

If we introduce the assumption os a discrete space-time [3] we must have
\[t=n\tau \ ,\ \ \ \vec{x}=\vec{j}\varepsilon \ ,\ \ \ n,{j}_{1},{j}_{2},{j}_{3}\in Z\]
\begin{equation}
T=N\tau \ ,\ \ \ \lambda =M\varepsilon \ ,\ \ \ N,M\in Z
\end{equation}
\[w={\frac{2\pi }{N\tau }}\ ,\ \ \ k={\frac{2\pi }{M\varepsilon }}\ ,\ \ \
{\frac{1}{N}},{\frac{1}{M}}\in Q\]
where $\varepsilon, \tau$ are the fundamental length and time.

From these quantities one constructs the discrete wave functions (for simplicity we use
only one spacial coordinate):
\begin{equation}
\psi \left({x,t}\right)=\exp\ \left\{{2\pi
i\left({{\frac{n}{N}}-{\frac{j}{M}}}\right)}\right\}
\end{equation}
wich is periodic in $n,j$ with period $N$ and wave length $M$.

We introduce an other wave function
\begin{equation}
\psi \left({x,t}\right)={\left({{\frac{1+{\frac{1}{2}}i{\frac{2\pi
}{N}}}{1-{\frac{1}{2}}i{\frac{2\pi
}{N}}}}}\right)}^{n}{\left({{\frac{1-{\frac{1}{2}}i{\frac{2\pi
}{M}}}{1+{\frac{1}{2}}i{\frac{2\pi }{M}}}}}\right)}^{j}
\end{equation}

This is a hot periodic function in $n$ or $j$, but is quasi-periodic in the sense that in
the limit $n\rightarrow \infty \ ,\ j\rightarrow \infty \ ,\ \tau \rightarrow 0\ ,\
\varepsilon \rightarrow 0\ ,\ n\tau =t\ ,\ j\varepsilon =x$,
\[\psi \left({x,t}\right)\rightarrow \exp\ \ i2\pi \left({wt-kx}\right).\]
which is periodic in $t$ and $x$

The arguments leading to de Broglie relations are translated into the discrete language.
The integral Lorentz transformations are factorized with the help of Kac generators [4]
\[{S}_{1}=\left({\begin{array}{cccc}1&0&0&0 \\
0&0&1&0\\
0&1&0&0\\
0&0&0&1\end{array}}\right),\ \ {S}_{2}=\left({\begin{array}{cccc}1&0&0&0\\
0&1&0&0\\
0&0&0&1\\
0&0&1&0\end{array}}\right)\]
\[{S}_{3}=\left({\begin{array}{cccc}1&0&0&0\\
0&1&0&0\\
0&0&1&0\\
0&0&0&-1\end{array}}\right),\ \ {S}_{4}=\left({\begin{array}{cccc}2&1&1&1\\
-1&0&-1&-1\\
-1&-1&0&1\\
-1&-1&-1&0\end{array}}\right)\]
in such a way that an element of the complete Lorentz group

\[L={P}_{1}^{\alpha }{P}_{2}^{\beta }{P}_{3}^{\gamma }{S}_{4}\ \ {P}_{1}^{\delta
}{P}_{2}^{\xi }{P}_{3}^{\eta }{S}_{4}\ ...\ {S}_{4}{S}_{1}^{\rho }{S}_{2}^{\sigma
}{S}_{3}^{\tau }\]
where ${P}_{1}={S}_{1}{S}_{2}{S}_{3}{S}_{2}{S}_{1}\ ,\ {P}_{2}={S}_{2}{S}_{3}{S}_{2}\ ,\
{P}_{3}={S}_{3}\ ;\ \alpha ,\beta ,\gamma ,\delta \xi \eta ,\rho ,\sigma ,\tau =0,1.$

The energy and the relativistic momentum are written
\begin{equation}
E={\frac{{m}_{0}{c}^{2}\left({c\Delta t}\right)}{{\left\{{{\left({c\Delta
t}\right)}^{2}-{\left({\Delta x}\right)}^{2}}\right\}}^{1/2}}}\ ,\ \ \
P={\frac{{m}_{0}c\Delta x}{{\left\{{{\left({c\Delta t}\right)}^{2}-{\left({\Delta
x}\right)}^{2}}\right\}}^{1/2}}}
\end{equation}
hence
\begin{equation}
{\frac{p{c}^{2}}{E}}={\frac{\Delta x}{\Delta t}}=u
\end{equation}

As in the continuous case, if ${v}_{\varphi }={c}^{2}/u$ $(E,p)$ transform in the same way
as $(w,k)$, hence 
\begin{equation}
E=\hbar w\ \ \ ,\ \ \ p=\hbar k
\end{equation}

The identification of $v_g=u$ is made by the superposition of two wave functions of slightly
different wave length and period [5]
\[\psi \left({x,t}\right)=\cos2\pi \left({{\frac{t}{T}}-{\frac{x}{\lambda
}}}\right)+\cos2\pi \left({{\frac{t}{T'}}-{\frac{x}{\lambda '}}}\right)\]
\[=2\cos\pi \left\{{t\left({{\frac{1}{T}}-{\frac{1}{T'}}}\right)-x\left({{\frac{1}{\lambda
}}-{\frac{1}{\lambda '}}}\right)}\right\}\cos\pi
\left\{{t\left({{\frac{1}{T}}+{\frac{1}{T'}}}\right)-x\left({{\frac{1}{\lambda
}}+{\frac{1}{\lambda '}}}\right)}\right\}\]

To this wave it corresponds a phase velocity
\begin{equation}
{v}_{\varphi }={\frac{{\frac{1}{T}}+{\frac{1}{T'}}}{{\frac{1}{\lambda }}+{\frac{1}{\lambda
'}}}}={\frac{\tilde{\Delta }w}{\tilde{\Delta }k}}
\end{equation}
and a group velocity
\begin{equation}
{v}_{g}={\frac{{\frac{1}{T}}-{\frac{1}{T'}}}{{\frac{1}{\lambda }}-{\frac{1}{\lambda
'}}}}={\frac{\Delta w}{\Delta k}}
\end{equation}
where $\Delta$ and $\tilde{\Delta}$ are the difference and average operator

From (16) (17) and (18) we have
\begin{equation}
{v}_{\varphi }={\frac{\tilde{\Delta }w}{\tilde{\Delta }k}}={\frac{\tilde{\Delta
}E}{\tilde{\Delta }p}}={\frac{{c}^{2}}{u}}
\end{equation}
\begin{equation}
{v}_{g}={\frac{\Delta w}{\Delta k}}={\frac{\Delta E}{\Delta p}}=u
\end{equation}

To prove the last equality in (22) we take the expression
${E}^{2}/{c}^{2}-{p}^{2}={m}_{0}^{2}{c}^{2}$ and apply the total
difference to both sides. From the definition $\Delta f\equiv f\left({x+\Delta
x}\right)-f\left({x}\right)$ we get $\Delta \left({{E}^{2}}\right)={\left({E+\Delta
E}\right)}^{2}-{E}^{2}$ and $\Delta \left({{p}^{2}}\right)={\left({p+\Delta
p}\right)}^{2}-{p}^{2}$. Therefore 
\begin{equation}
\left\{{2E\Delta E+{\left({\Delta E}\right)}^{2}}\right\}/{c}^{2}-2\vec{p}\Delta
\vec{p}-{\left({\Delta \vec{p}}\right)}^{2}=0
\end{equation}

The difference of momentum in two consecutive events of the particle $\Delta {p}_{\mu }$ is
also a 4-vector. In order to calculate the invariant ${\left({\Delta
E}\right)}^{2}/{c}^{2}-{\left({\Delta p}\right)}^{2}$ we take an inertial system such
that $\Delta {p}=0$ which means that the momentum is the same for two consecutive events,
and the velocity and energy are the same. Hence $\Delta E=0$, so that in an arbitrary
inertial frame.

\[{\left({\Delta E}\right)}^{2}/{c}^{2}-{\left({\Delta p}\right)}^{2}=0\]

Inserting this result in (23) and using (21) we obtain
\begin{equation}
\Delta E=u\Delta p
\end{equation}

To prove the last equality of (21) we apply the identity $\Delta
\left({{f}_{g}}\right)=\Delta f\ \tilde{\Delta }g+\tilde{\Delta }f\ \Delta g$ to the
expression ${E}^{2}/{c}^{2}-{p}^{2}={m}_{0}^{2}{c}^{2},$ as in the continuous case. 
 
\section{Physical consequences}

If we accept the assumption of a discrete space and time as a consequence of the
interaction of fundamental entities [6] we may conceive a vibration on this network,
similar to the waves propagating on a discrete string. The plane waves satisfy the
properties of section 2 therefore, we can talk of phase velocity and group velocity of
the packet. Those properties can be associated to a particle whose structure is attached
to the wave packet, but with experimental data given by the Einstein-de Broglie relations.

In particular we have the following physical consequences:

i) the frequency and the wave number are discrete, because the period and
wave length are integral multiple of fundamental time and length

ii) the energy and relativistic momentum are discrete due to the Einstein-de
Broglie relations
\begin{equation}
{\frac{{m}_{0}{c}^{2}c\Delta t}{{\left\{{{\left({c\Delta t}\right)}^{2}-{\left({\Delta
x}\right)}^{2}}\right\}}^{1/2}}}={\frac{h}{N\tau }}\ \ \ ,\ \ \ N\ integer
\end{equation}

\begin{equation}
{\frac{{m}_{0}c\Delta x}{{\left\{{{\left({c\Delta t}\right)}^{2}-{\left({\Delta
x}\right)}^{2}}\right\}}^{1/2}}}={\frac{h}{M\varepsilon }}\ \ \ ,\ \ \ M\ integer
\end{equation}

iii) in the rest system
\[{m}_{0}={\frac{h}{{c}^{2}}}{\frac{1}{N\tau }}\]
we have a discrete mass spectrum.

iv) the wave equation on the lattice reads [7]:
\[\left({-{\frac{1}{{c}^{2}}}{\frac{1}{{\tau }^{2}}}{\Delta }_{n}{\nabla
}_{n}{\tilde{\Delta }}_{j}{\tilde{\nabla }}_{j}+{\frac{1}{{\varepsilon }^{2}}}{\Delta
}_{j}{\nabla }_{j}{\tilde{\Delta }}_{n}{\tilde{\nabla }}_{n}}\right)\psi
\left({x,t}\right)={\frac{{m}_{0}^{2}{c}^{2}}{{\hbar }^{2}}}{\tilde{\Delta
}}_{j}{\tilde{\nabla }}_{j}{\tilde{\Delta }}_{n}{\tilde{\nabla }}_{n}\psi
\left({x,t}\right)\] where the solutions are given by (14) or (15) provided the dispersion
relations are satisfied
\[{\frac{1}{{c}^{2}}}{\frac{1}{{\tau }^{2}}}{\tan}^{2}{\frac{\pi
}{N}}-{\frac{4}{{\varepsilon }^{2}}}{\tan}^{2}{\frac{\pi
}{M}}={\frac{{m}_{0}^{2}{c}^{2}}{{\hbar }^{2}}}\]
in the first case and 
\[{\frac{1}{{c}^{2}}}{\left({{\frac{1}{N\tau
}}}\right)}^{2}-{\left({{\frac{1}{M\varepsilon
}}}\right)}^{2}={\frac{{m}_{0}^{2}{c}^{2}}{{h}^{2}}}\] in the second case.

\section*{Acknowledgments} This work has been partially supported by D.G.I.C.Y.T. (grant
PB94-1318)

\end{document}